\documentstyle[preprint,aps]{revtex}
\begin{document}
\preprint{
SUNY-RHI-95-8, in print Phys. Lett. {\bf B} Sept. 1995
}
\normalsize
\title { Thermal and Hadrochemical Equilibration in Nucleus-Nucleus Collisions
         at the SPS}
\author{P. Braun-Munzinger, J. Stachel, J. P. Wessels, and
N. Xu$^*$}
\address {\em Department of Physics \\
       State University of New York at Stony Brook \\
       Stony Brook, \, New York 11794 -- 3800 }
\date{Sep. 22, 1995}
\maketitle

\begin{abstract}
The currently available set of hadron abundances at the SPS for central
S + Au(W,Pb) collisions is compared to predictions from a scenario
assuming local thermal and hadrochemical equilibrium. The data are
consistent with a freeze-out temperature T = 160 - 170 MeV. Spectra are
consistent with this temperature range and a moderate transverse
expansion. The freeze-out points at the AGS and SPS are found to be
close to the phase boundary between a hadron gas and an ideal
quark-gluon phase.
\end{abstract}
\narrowtext

Studies of ultra-relativistic nucleus-nucleus collisions in fixed target
experiments at the BNL AGS and CERN SPS investigate hadronic matter at
extreme density. Nuclei at the AGS are found to stop each other
completely in the c.m. frame, while at the SPS an onset of transparency
is observed for S+A collisions \cite{stopping}. It has been predicted
\cite{hung} that the longest lived systems with high density are
produced somewhere between AGS and SPS energies. Hadronic cascade models
yield densities in the interior of the colliding nuclei up to ten times
normal nuclear matter density \cite{rqmd}. An intriguing possibility is
that the phase boundary to quark matter is crossed in these
collisions. For zero net baryon density the transition temperature
resulting from numerical simulations of QCD on a lattice is 150$\pm$10
MeV \cite{lattice}. A first order transition can be constructed for
finite baryon density or non-zero up/down quark chemical potential
between an ideal hadron gas and an ideal quark-gluon gas and the result
will be shown below. We will first address the question to what extent
one is allowed to actually talk about a phase in the thermodynamic
sense. The question is whether the time scale is long enough and/or the
collisions are frequent enough for the system to equilibrate before it
breaks up (``freezes out'') into the final state hadrons. Present data
are now detailed enough to address whether the system is in equilibrium
at freeze-out.

Early data on global observables were seen \cite{jspbm} as indicative of
a system with a temperature in the vicinity of 150 MeV, in local
equilibrium but expanding longitudinally.  Since then an extensive set
of data on hadronic abundances and spectral distributions has become
available for silicon and sulphur induced reactions and the hypothesis
of thermal as well as hadrochemical equilibrium has been addressed by
several authors
\cite{cley_ags,cley_cern,raf_ags,raf_sps,heinz,thermal,panagio}.  In
particular, in \cite{heinz} an impressive and complete survey is given
of theoretical techiques and their applicability to interpret data.  For
the AGS we have shown recently \cite{thermal} that the complete set of
hadron abundances is consistent with a system in equilibrium at a
temperature in the range 120 - 140 MeV, a baryon chemical potential of
540 MeV, and strangeness in equilibrium with up/down flavors. The system
appears to be in local equilibrium with an overall longitudinal and
transverse expansion with average velocities of $\langle \beta_l
\rangle$ = 0.52 and $\langle \beta_t \rangle$ = 0.39 - 0.33. Here we
will use the same approach to test whether data at SPS energies are
consistent with hadrochemical equilibrium as well.

We use for the present analysis the complete set of data now available
from the different experiments. Previous studies often concentrated
largely on the abundances of strange and multiply strange baryons, in
part by choice and in part because some of the relevant nonstrange
hadron and meson abundances were not available at the time. This
approach yields significant discrepancies between model predictions and
data when the most abundantly produced particles, pions, are considered,
as was already noted by Redlich at al. \cite{cley_cern}. Further, we
will use data integrated over the maximum available range of rapidity
and transverse momentum since flow effects can severely distort relative
hadron abundances at a given rapidity and transverse momentum. Similar
to \cite{cley_ags,cley_cern} we will investigate only the case of
complete strangeness saturation instead of allowing an additional free
parameter to govern abundances of strange hadrons as in
\cite{raf_ags,raf_sps,heinz,panagio}. In addition to comparing relative
hadron abundances to the data we will also compare the absolute pion
density from the thermal model to an estimate obtained from the
experimental data.

The basic assumption of our thermal model is that in every local
restframe the system is described by a grand canonical ensemble of
fermions and bosons in equilibrium at (freeze-out) temperature $T$. For
an infinite volume the particle number densities are given as integrals
over particle momentum $p$:
\begin{equation}
 \rho_i^0 = \frac{g_i}{2\pi^2} \int_{0}^{\infty} \frac{p^2 dp}{{\rm
exp}[(E_i-\mu_b B_i - \mu_s S_i)/T] \pm 1} \label{eq:dens}
\end{equation}
where $g_i$ is the spin-isospin degeneracy of particle {\it i}, $E_i$,
$B_i$ and $S_i$ are its total energy in the local rest frame, baryon
number and strangeness, and $\mu_b$ and $\mu_s$ are the baryon and
strangeness chemical potentials (unless otherwise noted $\hbar$
=c=1). Energy density $\epsilon_i^0$, pressure $P_i^0$, and
entropy density $\sigma_i^0$ for a given species are obtained by
evaluating equation~\ref{eq:dens} above with an additional factor of
$E_i$, $p^2/(3E_i)$, and $(p^2/(3E_i)-\mu_i+E_i)/T$ in the integrand,
respectively.

For a system of finite size the integrand in
equation (1) has to be multiplied  by a correction
factor \cite{finvol}. For an estimate of this correction we assume a spherical
volume
with radius $R$ giving a correction factor
\begin{equation}
  f = 1 - \frac{3\pi}{4 pR} + \frac{1}{(pR)^2}.  \label{eq:finvol}
\end{equation}
To approximately account for the volume taken up by baryons we apply an
excluded volume correction to the partition function,
\begin{equation}
 {\rm ln} Z_i = \frac{{\rm ln} Z_i^0}{1 + \sum_{j} V_j \rho_j^0}
\label{eq:exvol}
\end{equation}
where $V_j$ is the volume occupied by an individual baryon and the sum
extends over all baryons; we use a sharp sphere volume with radius
0.8 fm for all baryons. The excluded volume correction for the pressure
takes the same form as eq.~\ref{eq:exvol}; to correct the entropy,
particle, and energy densities the appropriate derivatives of the correction
factor with respect to temperature and/or chemical potential have to be
taken into account. This simple correction is appropriate for $\sum_{j}
V_j \rho_j^0 \leq 0.5$, valid for the area of interest in this paper.
More general (and much more involved) procedures are discussed in
\cite{prak}. The finite size and excluded volume corrections are
sizeable and affect the absolute densities and pressures, but ratios of
particle yields or quantities such as the entropy/baryon are affected to
a lesser extent.

For the comparison to SPS heavy ion data we include in the thermal model
all known \cite{databook} baryons and mesons up to a mass of 2 and 1.5
GeV, respectively. We have checked that for $T \leq 180$ MeV higher mass
mesons and baryons do not play an important role and lead to corrections
of less than 5~\% in the particle densities.  The correction for feeding
and decay is performed as in \cite{thermal} using all known branching
ratios \cite{databook} and symmetry and phase space arguments for
unknown branching ratios. Unless an explicit number is quoted by the
experiment, we assume a 50~\% particle identification efficiency
following weak decays.

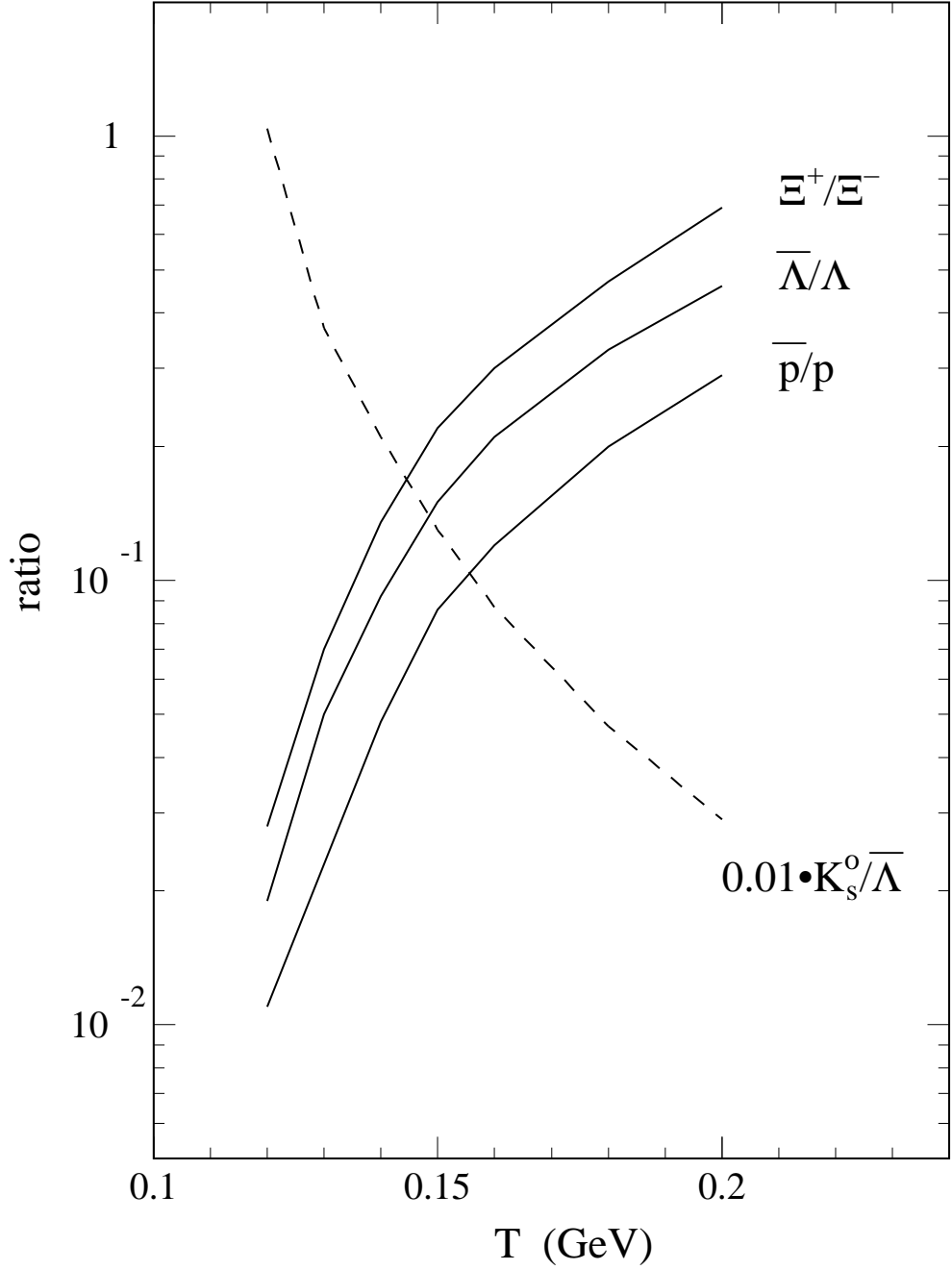
\begin{figure}
\caption{Ratios of baryon abundances that show a strong temperature
dependence for a fixed pion to nucleon ratio as indicated in Table\
\ref{ratios}. See text for details.}
\label{ratio}
\end{figure}

The starting point for our thermal model calculations is to determine,
for each temperature, that baryon chemical potential with which are best
described experimental data reflecting the proton to pion ratio (first
four rows in Table\ \ref{ratios}). As in \cite{thermal} the strangeness
chemical potential is fixed by the strangeness neutrality
condition. Best overall agreement with all data
is obtained with T = 160 - 170 MeV, as shown in Table\ \ref{ratios}
where all currently available experimental data on particle ratios
measured in central S + Au(W,Pb) collisions are compared to predictions
of the thermal model. A graphic illustration of what drives the
temperature in our freeze-out analysis can be seen in Figure\
\ref{ratio}, where predictions for those particle ratios that show,
after our choice of $\mu_b$, the largest temperature sensitivity are
plotted {\it vs} T. Typical changes are about a factor 30 over the
temperature range considered and good overall agreement with the data
for these temperature sensitive ratios is obtained for T = 160 - 170 MeV
and corresponding baryon chemical potentials of 170 - 180 MeV,
respectively. Since we assume strangeness equilibration all other
particle ratios, including those for strange and multi-strange hadrons,
are then fixed. Inspection of Table\ \ref{ratios} shows that with these
two parameters one can obtain a surprisingly good description of the
available experimental data.

Most measured particle ratios are reproduced well within the errors,
especially keeping in mind systematic errors and systematic acceptance
effects as displayed e.g. in Table\ \ref{ratios} by data for the same
ratio from different experiments. Of the 27 measured ratios spanning a
range of a factor 400 just three differ beyond statistical errors by
40-50~\% and the only more serious discrepancy is the ratio
$(\Omega^++\Omega^-)/(\Xi^++\Xi^-)$ which appears to be about a factor 2
different. Note, however, that there is no systematic indication that a
separate parameter is needed to control strangeness abundance. At the
same time, there are significant differences (up to a factor of four) in
the ratios for the heavy ion reaction considered here as compared to
nucleon nucleon data \cite{gazd2}.

The freeze-out temperature resulting from our analysis is considerably
lower than that proposed by \cite{raf_sps} and \cite{heinz} to describe
data. Although in the latter a range of temperatures is considered, the
finally proposed chemical freeze-out temperature is 190 MeV causing, as
the authors note, a significant underprediction in the pion abundance as
compared to data.  The lower temperature is reassuring, because
temperatures around 200 MeV will lead to absolute pion densities of
0.6-0.7/fm$^3$ after excluded volume correction, {\it i.e.}  even pions
start to overlap significantly (as noted also in
\cite{heinz}). Furthermore, such a high freeze-out temperature would
imply that freeze-out takes place well in the quark-gluon plasma region
of the phase diagram (see below). In fact, experimental data on two-pion
interferometry can be used to obtain an estimate of the pion density at
freeze-out. We begin by noting that experimental $\pi^+\pi^+$ and
$\pi^-\pi^-$ correlations are rather well described by a simulation
using the cascade code RQMD \cite{na44-hbt}. Next we inspect the
space-time history of particles in RQMD. Pions freeze out over an
extended time in RQMD during which the source is expanding. To give a
typical volume we quote here the size of the system at the average pion
freeze-out time t = 14 fm/c. At mid-rapidity the distribution of
positions transverse to the beam is Gaussian with standard deviation
$\sigma_x = \sigma_y$ = 4.0 fm, the longitudinal distribution for low
transverse momenta ($p_t \leq$ 0.1 GeV/c) has a standard deviation of
$\sigma_z$ = 7.7 fm. This corresponds to a volume $V =
(2\pi)^{3/2}\sigma_x\sigma_y\sigma_z$ = 1940 fm$^3$. The rapidity
integrated multiplicity for negatively charged particles for central
S+Au collisions has recently been reported by NA35 \cite{gazd} and from
their data we estimate that the total pion multiplicity in these
collisions is about 510. Using the volume estimate just given one
obtains a typical pion density at freeze-out $\rho_{\pi} \approx
0.27/{\rm fm^3}$. This is rather close to the value of 0.30/fm$^3$
obtained in the thermal model for T = 160 MeV and the above
parameters. Other volume estimates, e.g. obtained by using the time
averaged variances or the variances of the time integrated
distributions, are larger, leading to lower freeze-out density
estimates.

The entropy at freeze-out, when evaluated in the thermal model, is found
to be large. For T = 160 and 170 MeV we obtain for the entropy per net
baryon, which is equivalent to the entropy produced per incident
nucleon, values of S/(B-\=B) = 45.4 and 36.7, respectively.

The strange chemical potentials obtained for $T$ = 160 (170) MeV and
$\mu_b$ = 170 (180) MeV are $\mu_s$ = 38.0 (47.0) MeV. On the quark
level this implies a strange quark chemical potential of $\mu_{qs} =
\frac{1}{3} \mu_b - \mu_s$ = 18.6 (13.0) MeV, much smaller than the
strange quark mass. Obviously, for every temperature $T$ there is one
value of $\mu_b$ which, together with strangeness conservation, will
yield $\mu_{qs}$ = 0. In the present case the pion to nucleon ratio
happens to be such that we are close to this situation. This situation
arises in a purely hadronic picture and is not related to whether or not
the system is in the quark-gluon plasma phase at freeze-out.

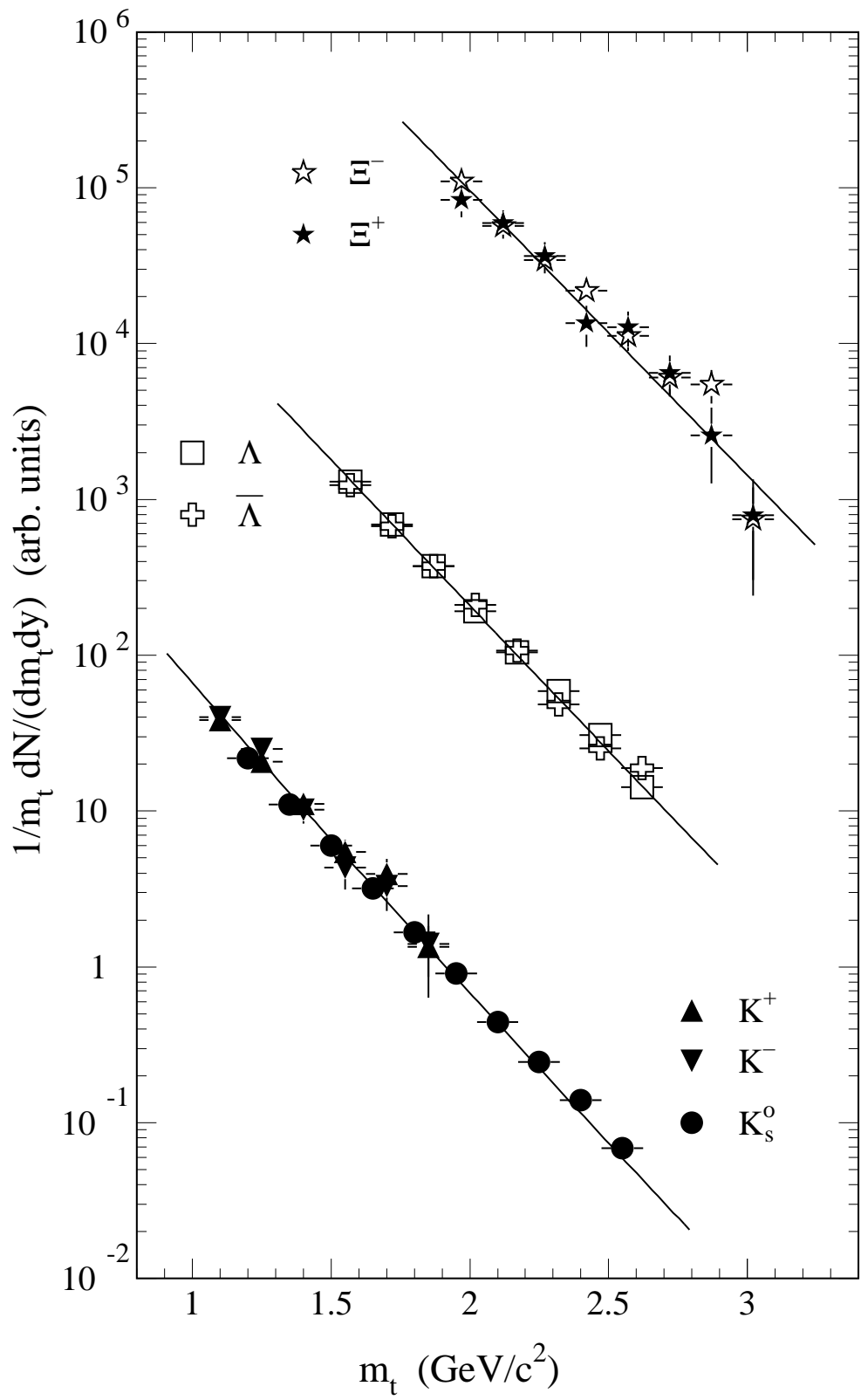
\begin{figure}
\caption{Kaon, Lambda and Cascade spectra from WA85 [28,24]
compared to thermal model calculations with T = 160 MeV and an average
transverse expansion velocity of $\langle \beta_t \rangle$ = 0.27 (solid
lines).}
\label{spectra}
\end{figure}

In order to test whether the transverse momentum spectra of the various
hadrons are consistent with a temperature of 160 MeV and one common
transverse flow velocity we use, as in our analysis for the AGS
\cite{thermal}, the formalism developed in \cite{ekkard}. To minimize
systematic uncertainties, we choose spectra of different particle
species measured in the same experiment and close to mid-rapidity and we
have made the comparison for three sets: pion, kaon, proton and deuteron
spectra from NA44 \cite{dodd}, where a similar fit is also shown;
spectra of kaons, lambdas and cascades from WA85 \cite{abat,bari} shown
together with the fit in Fig.\ \ref{spectra}; the $\eta$ to pion ratio
from WA80 \cite{santo}.  All spectra are consistent with a temperature
of 160 MeV and, for a linear dependence of the flow velocity on the
radius of the system ($\alpha$ = 1), an average transverse expansion
velocity of $\langle \beta_t \rangle$ = 0.27. Consistent results were
reported by \cite{ekkard} for spectra from S+S collisions. This value is
somewhat smaller than the range of 0.33 - 0.39 found at AGS energies
\cite{thermal}.

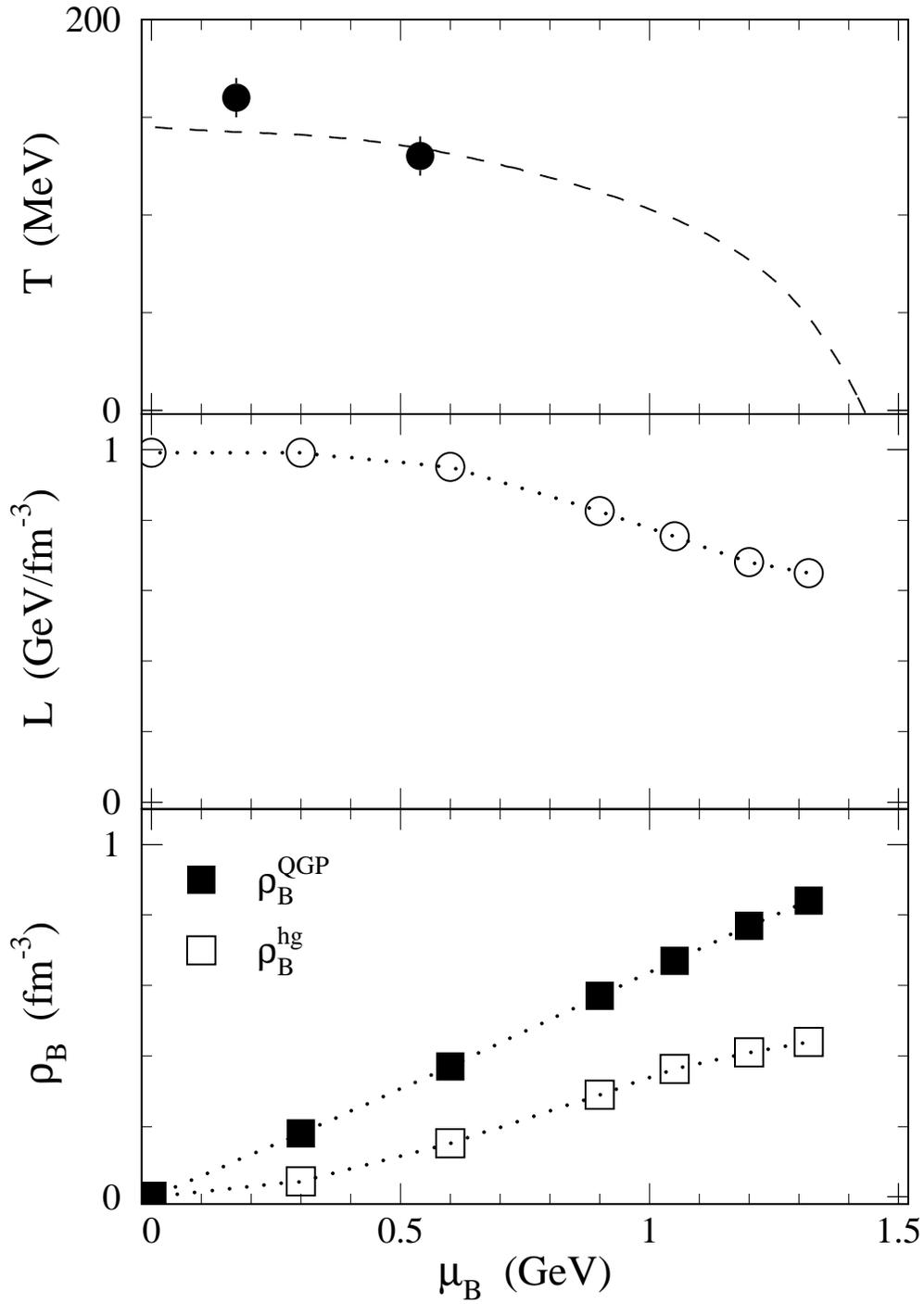
\begin{figure}
\caption{Phase boundary between a hadron gas and a quark-gluon plasma
(top) as function of temperature and baryon chemical potential together
with the freeze-out points for Si(S) + Au(W,Pb) collisions at AGS [11]
and SPS (present paper) energies. Latent heat of the
phase transition (middle) and baryon density in the hadron and
quark-gluon phase at the phase boundary (bottom).}
\label{phase}
\end{figure}

To put the above determined freeze-out temperature and baryon chemical
potential into perspective we have calculated the phase boundary between
the hadron resonance gas and the quark-gluon plasma by equating the
chemical potentials and the pressure in the hadronic phase, with those
of an idealized phase of massless u,d quarks, s quarks with mass of 150
MeV, gluons, and a bag constant of B = 262 MeV/fm$^3$. The resulting
phase diagram is shown in Figure\ \ref{phase} along with the latent heat
and baryon densities at the transition line. Interactions among the
hadrons, which are neglected in our approach, are not expected to change
this phase boundary by much \cite{prak} except possibly at high baryon
density ($\mu_b > 1$ GeV), {\it i.e.} far away from the freeze-out
region. The freeze-out points determined from the present analysis and
from that for AGS data \cite{thermal} are shown by the filled circles in
this diagram. Note that, with this scenario, the system at freeze-out,
i.e. {\bf after } expansion and cooling, is close to the phase boundary
at both AGS and CERN energies.

\vspace{0.2cm}

We would like to thank M. Prakash and E. Shuryak for enlightening
discussions.  This work was supported in part by the National Science
Foundation. One of us (J. P. W.) is supported by the A. v. Humboldt
Foundation as a F. Lynen fellow.

\vspace{0.2cm}

\noindent $^*$ present address: P-25 MS D456, LANL, Los Alamos, NM 87545.

\begin{table}
\caption{Particle ratios calculated in a thermal model for temperatures
of 160 and 170 MeV, baryon chemical potential $\mu_b$ of 170 and 180 MeV
and strangeness chemical potential $\mu_s$ of 38.0 and 47.0 MeV, in
comparison to experimental data (with statistical errors in parentheses)
for central collisions of 200 A GeV/c S + Au(W,Pb). For experimental data
a $p_t$ range is quoted when the lower limit is significantly larger
than zero.}

\label{ratios}
\newpage
\begin{tabular}{||c|ll|llcc||}
\multicolumn{1}{||c|}{Particles} & \multicolumn{2}{c|}{Thermal Model} &
\multicolumn{4}{c||}{Experimental Data}\\
 & \multicolumn{2}{c|}{T(MeV)} & \multicolumn{2}{c}{ } &
\multicolumn{2}{c||}{ }\\
 & 160 & 170 & exp. ratio & ref. & y & $p_t$ \\ \hline
p/$\pi^+$ & 0.17 & 0.19 & 0.18(3) &
NA44\cite{murray},NA35\cite{gazd} & 2.6-2.8 &\\
pos-neg/neg & 0.18 & 0.21 & 0.15(1) & NA35\cite{mitch} & 4-5.8 & \\
p-\=p/neg & 0.13 & 0.14 & 0.15(2) & NA35\cite{roehrich,gazd} & 3.2-5.4 & \\
pos-neg/pos+neg & 0.084 & 0.094 & 0.088(7) & EMU05\cite{emu05} & 2.3-3 & \\
d/p & 0.014 & 0.017 & 0.015(2) & NA44\cite{simon} & 1.8-2.5 &\\
\=p/p & 0.13 & 0.14 & 0.12(2) & NA44\cite{jacak} & 2.65-2.95 & \\
\=p/$\pi^-$ & 0.022 & 0.027 & 0.024(9) &
NA44\cite{murray},NA35\cite{gazd} & 2.6-2.8 &\\
$\eta$/$\pi^0$ & 0.12 & 0.12 & 0.15(2) & WA80\cite{santo} & 2.1-2.9 &\\
$\phi$/($\rho+\omega$) & 0.11 & 0.12 & 0.080(20) & Helios3\cite{massera} &
$\geq$3.5 & \\
\hline
K$^+/\pi^+$ & 0.21 & 0.22 & & & &\\
K$^+$+K$^-$/K$^0_s$ & 1.05 & 1.06 & 1.07(3) & WA85\cite{bari} &
2.5-3.0 & 1-2\\
K$^+$/K$^-$ & 1.46 & 1.53 & 1.67(15) & WA85\cite{bari} & 2.3-3.0 & $>$0.9 \\
K$^0_s$/$\Lambda$ & 1.74 & 1.50 & 1.4(1) & WA85\cite{bari} & 2.5-3.0 &
1-2.5 \\
 & 1.57$^a$ & 1.36$^a$ & 0.88(10) & NA35\cite{alber} & 3.5-5.5 & \\
K$^0_s$/$\bar{\Lambda}$ & 8.5 & 6.6 & 6.4(4) & WA85\cite{bari} & 2.5-3.0 &
1-2.5 \\
 & 7.3$^a$ & 5.7$^a$ & 4.6(10) & NA35\cite{alber} & 3.5-5.5 & \\
\hline
$\Lambda$/(p-\=p) & 0.67 & 0.69 & 0.45(4) & NA35\cite{roehrich} &
3.25-5.25 & \\
$\bar{\Lambda}$/\=p & 0.38 & 0.41 & 0.80(30) & NA35\cite{guenther} &
3.25-5.0 & \\
$\bar{\Lambda}/\Lambda$ & 0.20$^b$ & 0.23$^b$ & 0.20(1) & WA85\cite{abat} &
2.3-3.0 & 1.2-3\\
 & & & 0.207(12) & NA36\cite{andersen} & 1.5-3.0 & 0.6-1.6\\
 & 0.22$^a$ & 0.24$^a$ & 0.19(4) & NA35\cite{alber} & 3.5-5.5 & \\
\hline
$\Xi^- / \Lambda$ & 0.12$^b$ & 0.12$^b$ & 0.095(6) & WA85\cite{abat} & 2.3-5.0
& 1.2-3\\
 & & & 0.066(13) & NA36\cite{andersen} & 1.5-2.5 & 0.8-1.8\\
$\Xi^+ / \bar{\Lambda}$ & 0.20$^b$ & 0.21$^b$ & 0.21(2) & WA85\cite{abat} &
2.3-3.0 & 1.2-3\\
 & & & 0.127(22) & NA36\cite{andersen} & 2.0-3.0 & 0.6-1.8\\
$\Xi^- / \Xi^+$ & 0.31 & 0.36 & 0.45(5) & WA85\cite{abat} & 2.3-3.0 & 1.2-3\\
 & & & 0.276(108) & NA36\cite{andersen} & 2.0-2.5 & 0.8-1.8\\
\hline
$(\Omega^++\Omega^-)/(\Xi^++\Xi^-)$ & 0.17 & 0.19 & 0.8(4) &
WA85\cite{bari} & 2.5-3.0 & $>$1.6\\
\hline
\=d/\=p & .0015 & .0018 & & & & \\
\end{tabular}
$^a$Reconstruction efficiency $\epsilon$ = 1 for particles from weak decays.\\

\vspace{-1.3cm}

$^b$ Yields of $\Lambda$,$\bar{\Lambda}$ corrected for feeding from $\Xi$.

\end{table}

\end{document}